\documentclass{PoS}
\usepackage[utf8]{inputenc}
\usepackage[T1]{fontenc}
\usepackage{amsmath}
\usepackage{graphicx}

\newcommand{\td}[1]{\, \mbox{d} #1 \,}

\newcommand{\fermi}{\textit{Fermi}-LAT}
\newcommand{\g}{\ensuremath{\gamma}}
\newcommand{\cta}{CTA\,102} 
\newcommand{\zred}{z_{\rm red}}

\newcommand{\logb}[1]{\ln{\left( #1 \right)}}
\newcommand{\p}{^{\prime}}

\newcommand{\E}[1]{\times 10^{#1}}

\title{CTA\,102 -- year over year receiving you}

\ShortTitle{CTA\,102 -- year over year receiving you}

\author{\speaker{Michael Zacharias}\\
        Ruhr Astroparticle and Plasma Physics Center (RAPP Center), Insitut f\"ur theoretische Physik IV, Ruhr-Universit\"at Bochum, D-44780 Bochum, Germany \\
	Centre for Space Research, North-West University, Potchefstroom 2520, South Africa\\
        E-mail: \email{mz@tp4.rub.de, mzacharias.phys@gmail.com}}

\author{Markus B\"ottcher\\
        Centre for Space Research, North-West University, Potchefstroom 2520, South Africa\\
        E-mail: \email{Markus.Bottcher@nwu.ac.za}}
\author{Felix Jankowsky\\
        Landessternwarte, Universit\"at Heidelberg, K\"onigstuhl, D-69117 Heidelberg, Germany\\
        E-mail: \email{jankowsky@lsw.uni-heidelberg.de}}
\author{Jean-Philippe Lenain\\
        Sorbonne Universit\'e, Universit\'e Paris Diderot, Sorbonne Paris Cit\'e, CNRS/IN2P3, Laboratoire de Physique Nucl\'eaire et de Hautes Energies, LPNHE, 4 Place Jussieu, F-75252 Paris, France\\
        E-mail: \email{jlenain@in2p3.fr}}
\author{Stefan J. Wagner\\
        Landessternwarte, Universit\"at Heidelberg, K\"onigstuhl, D-69117 Heidelberg, Germany\\
        E-mail: \email{swagner@lsw.uni-heidelberg.de}}
\author{Alicja Wierzcholska\\
        Institute of Nuclear Physics, Polish Academy of Sciences, PL-31342 Krakow, Poland\\
        Landessternwarte, Universit\"at Heidelberg, K\"onigstuhl, D-69117 Heidelberg, Germany\\
        E-mail: \email{a.wierzcholska@lsw.uni-heidelberg.de}}

\abstract{The FSRQ CTA 102 (z=1.032) has been tremendously active over the last few years. During its peak activity lasting several months in late 2016 and early 2017, the gamma-ray and optical fluxes rose by up to a factor 100 above the quiescence level. We have interpreted the peak activity as the ablation of a gas cloud by the relativistic jet, which can nicely account for the months-long lightcurve in 2016 and 2017. The peak activity was in the middle of a 2-year-long high-state, which was characterized by increased fluxes and increased rms variability compared to the previous low-states, and which was flanked by two bright flares. In this presentation, we put the cloud-ablation scenario into the broader context of the 2-year-long high-state.
}

\FullConference{High Energy Phenomena in Relativistic Outflows VII - HEPRO VII\\
		9-12 July 2019\\
		Facultat de Física, Universitat de Barcelona, Spain}

\begin{document}

\section{Introduction}

\cta\ was detected as a radio source in 1960 \cite{hr60}. It presented itself as a rather peculiar source, which culminated in \cta\ becoming the content of a pop song\footnote{The Byrds: ``C.T.A. 102'' (1967, Album: Younger than yesterday)}, which inspired the title of this contribution. While parts of \cta's peculiarities have been explained by the fact that it is a blazar, it is still able to surprise.

As many blazars \cite{z19}, \cta\ (located at a redshift of $\zred=1.037$) is a highly variably source. The variability originates in the relativistic jet, which is aligned with the line-of-sight. The relativistic aberration foreshortens the jet-intrinsic time scales and enhances the fluxes in the observer's frame making blazars ideal targets to study the physics of relativistic jets. Owing to the unpredictable nature of blazars, it is important to monitor many sources in as many energy bands as possible. Most notably, the continuous monitoring of the \g-ray sky with the {\it Fermi} satellite has proven to be an invaluable tool.

\begin{figure*}
\centering
\includegraphics[width=1.00\textwidth]{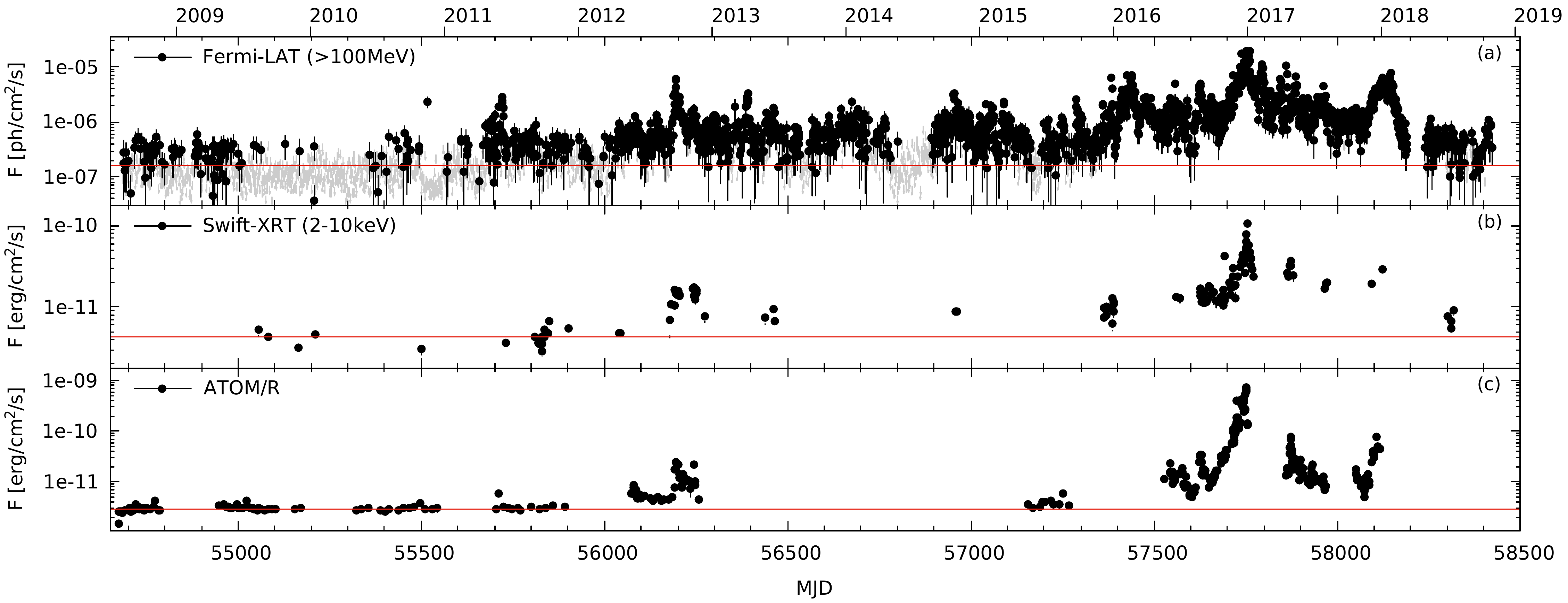}
\caption{Long-term light curves of CTA\,102. In all panels, the red line marks the average before 2012. Note the logarithmic scale on the y-axis.  
{\bf (a)} Daily HE \g-ray light curve observed with \fermi. The gray arrows mark upper limits.
{\bf (b)} {\it Swift}-XRT light curve for individual pointings.
{\bf (c)} ATOM R-band light curve for individual pointings.
Figure from \cite{zea19}.
}
\label{fig:ltlc}
\end{figure*} 
Observations with \fermi\ of \cta\ reveal a diverse and changing source (see panel (a) in Fig.~\ref{fig:ltlc}). Initially after the launch of {\it Fermi}, the source was \g-ray quiet with barely a detection on a daily scale. This changed in 2012 around a prominent flare \cite{lea16}, after which the source has become detectable on daily time scales. In early 2016 a strong flare marked another change in source behavior. Subsequently, the source remained on an elevated level a factor $\gtrsim 2$ above the previous level. In November 2016, fluxes began rising steadily over the course of two months culminating in an extreme high state with fluxes reaching more than $50$ times pre-November levels around new year 2017. At this time, \cta\ was among the brightest \g-ray sources in the sky despite its large distance from Earth. Over the next two months, the flux level fell back to pre-November levels and remained there until another strong outburst at the beginning of 2018 after which source fluxes seemingly fell back to pre-2016 levels. While the observations in the X-ray band with {\it Swift}-XRT (Fig.~\ref{fig:ltlc}(b)) and optical R-band with ATOM (Fig.~\ref{fig:ltlc}(c)) are not as detailed as the \fermi\ observations, one can deduce similar flux evolutions as in the \g-ray band. Most notably, the bright flare in 2016-2017 is also visible in these bands. In fact, the fluxes in the optical band during the flare in 2016-2017 increased by a factor $100$, while in the X-ray band fluxes increased by a factor $10$.

We have successfully reproduced the tremendous flux changes during the 2016-2017 flare by invoking a significant increase in particle number in the jet. These additional particles originate from a gas cloud that collides with the jet and which is gradually ablated causing the four-months-long flare. Another model \cite{rea17} explains the flare through a wobbling jet, where changes between the direction of motion and the line-of-sight cause changes in the Doppler boosting of the emitted radiation. While this model has the advantage that mild changes in the viewing angle already result in huge flux differences, the apparent chaotic motion of the jet remained unexplained.

\section{The cloud ablation model}

\begin{figure}
\centering
\includegraphics[width=0.35\textwidth]{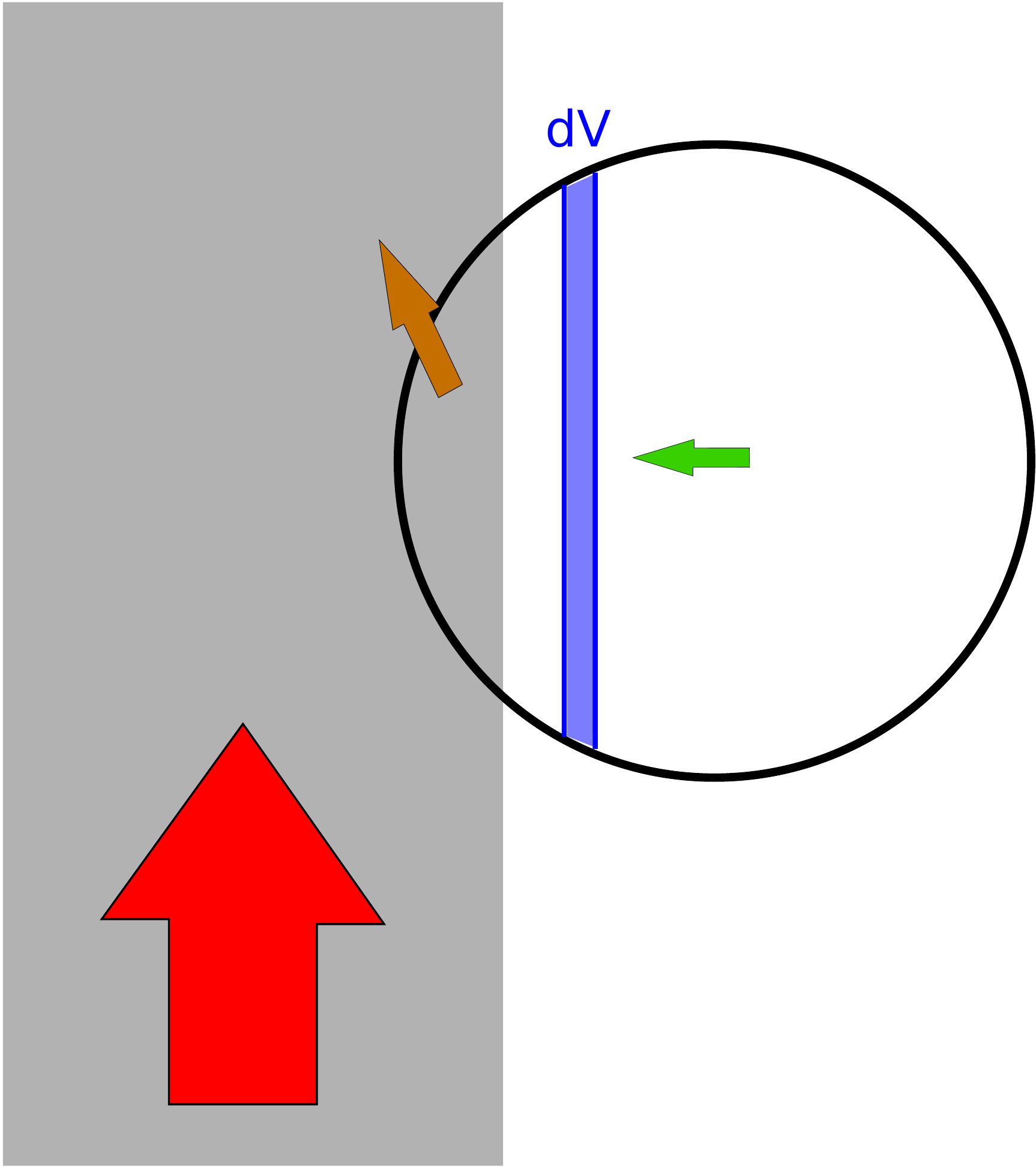}
\caption{Sketch of the ablation of a gas cloud. The gray area is a part of the jet with the red arrow depicting its ram pressure. The cloud is the black sphere with its velocity component orthogonal to the jet indicated by the green arrow. The brown arrow depicts the motion of the cloud material that has been ablated by the jet (parts of the cloud within the jet). The blue area is an example how the volume increment $\td{V}$ changes with time. Figure from \cite{zea19}.
}
\label{fig:cloud}
\end{figure} 

The injection of particles into the jet is a consequence of the time-dependent intrusion of the cloud, and the subsequent ablation of the cloud particles owing to the high ram pressure of the jet (c.f. Fig.~\ref{fig:cloud}). The details of the process depend strongly on the speed and size of the cloud.
The jet's ram pressure induces a shock in the cloud, which crosses the cloud with roughly the relativistic sound speed of $c/\sqrt{3}$, where $c$ is the speed of light. If the speed of the cloud is high, the cloud penetrates deep into the jet before the shock has crossed the cloud. In this case, the cloud material will be shocked in one instance resulting in a violent burst of particles and radiation \cite{abr10}. This provides the ingredients for fast flares lasting less than a day \cite{bea12,bba12}.

In the opposite scenario, the cloud takes a long time (maybe a few weeks) to penetrate the jet, because the cloud is large and/or slow. Therefore, a shock, which may take about 1 day to cross the cloud, mostly interacts with the subvolume of the cloud that has entered the jet during the time of the shock crossing. This results in a gradual ablation of the cloud \cite{zea17}. Simulations by \cite{pbb17} involving a jet-star interaction suggest that the ablation of the stellar atmosphere begins at the transition layer at the edge of the jet and proceeds while the star moves into the jet. The ablated material is subsequently mixed into the inner jet flow, where it is carried along and accelerated. Production of radiation \cite{tb19} may take place close to the interaction site through shock waves induced by the jet-cloud interaction or at a recollimation shock downstream in the jet. In any case, the relatively smooth injection of the cloud material into the jet flow leads to a gradual particle density enhancement resulting in a similar reaction of the radiative output.

Under the assumption that all particles of the cloud enter the jet, the injection of particles into the emission region of the jet yields the following dependency:

\begin{align}
 \frac{\td{N}}{\td{t}} \propto \logb{\frac{t_0^{2} + t_c^{2}}{t_0^{2}+(t_c-t)^2}} \label{eq:Ninj}
\end{align}
where $t_c= R\p_c/(\Gamma v\p_c)$ is related to the cloud radius $R\p_c$, and $t_0 = r\p_0/(\Gamma v\p_c)$ to the cloud's scale height $r\p_0$ \cite{zea17}. The time scales also depend on the speed of the cloud $v\p_c$ and the bulk Lorentz factor of the jet $\Gamma$ (as the injection takes place in the comoving frame of the jet, while the cloud parameters are determined in the galactic frame marked by primes). The scale height depends on the central density $n\p_0$ and the temperature $T\p$ of the cloud as $r\p_0\propto\sqrt{T\p/n\p_0}$. We have assumed that the cloud is isothermal and bound by its own gravity.

\section{Results}

\begin{figure*}
\begin{minipage}{0.49\textwidth}
\includegraphics[width=0.95\textwidth]{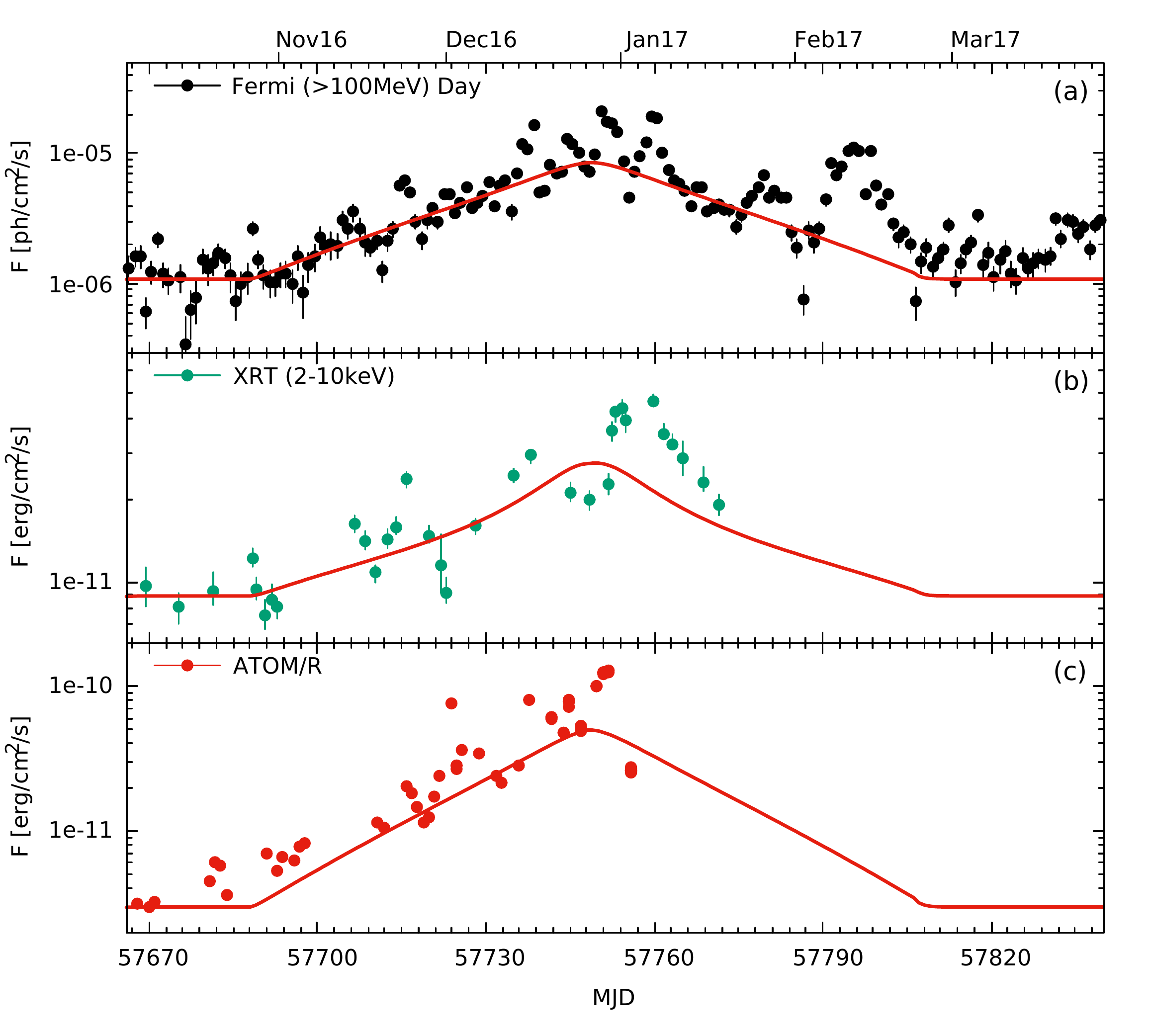}
\caption{Lightcurves and leptonic modeling (red lines) for the \g-ray band (a), X-ray band (b), and optical R band (c). Note the logarithmic y axis.
Figure from \cite{zea17}.
}
\label{fig:lep}
\end{minipage}
%
~
%
\begin{minipage}{0.49\textwidth}
\includegraphics[width=0.95\textwidth]{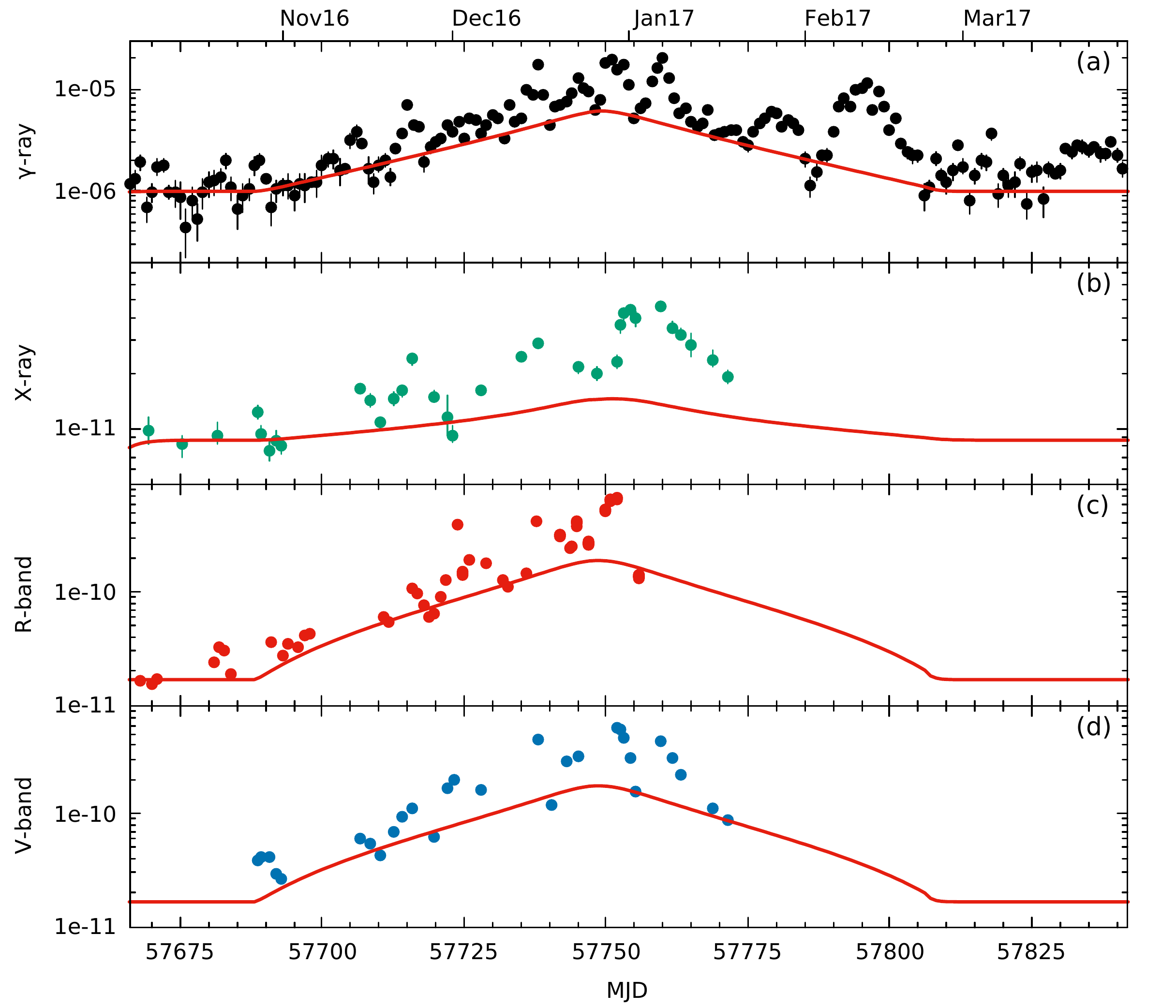}
\caption{Lightcurves and hadronic modeling (red lines) for the \g-ray band (a), X-ray band (b), optical R band (c), and optical V band (d). Note the logarithmic y axis.
Figure adapted from \cite{zea19}.
}
\label{fig:had}
\end{minipage}
\end{figure*} 
\begin{table}
\caption{Leptonic and hadronic model parameters. Primed parameters are in the galactic frame. Parameters below the first horizontal line mark the initial jet power (including all constituents) and ram pressure , while the parameters below the second horizontal line mark the variability parameters. Below the double horizontal line the inferred cloud parameters are given.}
\begin{tabular}{lccc}
Definition				& Symbol 				& Leptonic Model		& Hadronic Model \\
\hline
Doppler factor				& $\delta$				& $35\,$			& $35\,$ \\ 
Emission region distance		& $z$					& $6.50\times 10^{17}\,$cm	& $3.09\times 10^{18}\,$cm \\
Emission region radius			& $R$					& $2.5\times 10^{16}\,$cm	& $2.0\times 10^{16}\,$cm \\ 
Initial Magnetic field			& $B$					& $3.7\,$G			& $60\,$G \\ 
Proton injection luminosity		& $L_{\rm p,inj}$			& -				& $1.3\times 10^{44}\,$erg/s	\\
Minimum proton Lorentz factor		& $\gamma_{\rm p,min}$			& -				& $1.0\times 10^6\,$ \\ 
Initial maximum proton Lorentz factor	& $\gamma_{\rm p,max}$			& -				& $1.0\times 10^9\,$ \\ 
Proton spectral index			& $s_{\rm p}$				& -				& $2.4\,$ \\ 
Electron injection luminosity		& $L_{\rm e,inj}$			& $2.2\times 10^{43}\,$erg/s	& $3.2\times 10^{41}\,$erg/s	\\
Minimum electron Lorentz factor		& $\gamma_{\rm e,min}$			& $1.3\times 10^1\,$		& $2.0\times 10^2\,$ \\ 
Maximum electron Lorentz factor		& $\gamma_{\rm e,max}$			& $3.0\times 10^3\,$		& $3.0\times 10^3\,$ \\ 
Electron spectral index			& $s_{\rm e}$				& $2.4\,$			& $2.8\,$ \\ 
Escape time scaling			& $\eta_{\rm esc}$			& $10.0\,$			& $5.0\,$ \\ 
Acceleration to escape time ratio	& $\eta_{\rm acc}$			& $1.0\,$			& $30.0\,$ \\ 
Accretion disk luminosity		& $L_{\rm AD}\p$			& $3.75\times 10^{46}\,$erg/s 	& $3.75\times 10^{46}\,$erg/s\\ 
Radius of the BLR			& $R_{\rm BLR}\p$			& $6.7\times 10^{17}\,$cm	& $6.7\times 10^{17}\,$cm \\
Temperature of the BLR			& $T_{\rm BLR}\p$			& $1.0\times 10^4\,$K 		& $1.0\times 10^4\,$K \\
Luminosity of the BLR			& $L_{\rm BLR}\p$			& $4.14\times 10^{45}\,$erg/s	& $4.14\times 10^{45}\,$erg/s \\
Radius of the DT			& $R_{\rm DT}\p$			& $6.18\times 10^{18}\,$cm	& $6.18\times 10^{18}\,$cm \\
Temperature of the DT			& $T_{\rm DT}\p$			& $1.2\times 10^3\,$K		& $1.2\times 10^3\,$K \\
Luminosity of the DT			& $L_{\rm DT}\p$			& $7.0\times 10^{45}\,$erg/s 	& $7.0\times 10^{45}\,$erg/s \\
\hline
Initial jet power			& $P_{\rm jet}\p$			& $2.5\times 10^{46}\,$erg/s	& $7.0\times 10^{48}\,$erg/s \\
Initial jet ram pressure		& $p_{\rm ram}\p$			& $4.2\times 10^2\,$dyn/cm$^2$	& $3.1\times 10^3\,$dyn/cm$^2$ \\
\hline
Proton injection variability		& $\Delta{L}_{\rm p}$	 		& -				& $5.0\times 10^{43}\,$erg/s \\
Proton spectral index variability	& $\Delta{s}_{\rm p}$	 		& -				& $-0.3$  \\
Electron injection variability		& $\Delta{L}_{\rm e}$			& $9.5\times 10^{41}\,$erg/s	& $8.0\times 10^{41}\,$erg/s \\
Electron spectral index variation	& $\Delta{s}_{\rm e}$			& $-0.6\,$ 			& -	 \\
\hline
\hline
Cloud speed				& $v_{\rm c}\p$				& $5.1\E{8}\,$cm/s		& $1.9\E{8}\,$cm/s	 \\ 
Cloud radius				& $R_{\rm c}\p$				& $1.3\E{15}\,$cm		& $4.9\E{14}\,$cm	 \\ 
Cloud density				& $n_{\rm c}\p$				& $2.5\E{8}\,$cm$^{-3}$		& $1.1\E{7}\,$cm$^{-3}$	 \\ 
Cloud temperature			& $T_{\rm c}\p$				& $0.5\,$K			& $2.7\E{-3}\,$K	

\end{tabular}
\label{tab:param}
\end{table}

Using the particle injection function, Eq.~(\ref{eq:Ninj}), in leptonic and hadronic radiation codes \cite{db14,dbf15} provides the model lightcurves presented in Figs.~\ref{fig:lep} and \ref{fig:had}, respectively. In the leptonic code, synchrotron and inverse-Compton emission on synchrotron and broad-line regaion (BLR) target photons is most important, while the photon fields of the accretion disk and the dusty torus (DT) do not result in high inverse-Compton fluxes. In the hadronic code, all charged particles (electrons, protons, pions and muons) produce synchrotron emission, which provides the bulk of the electromagnetic emission. The parameters used to obtain the model lightcurves are given in Tab.~\ref{tab:param}. In this table, we also provide the initial total jet power and jet ram pressure. In the leptonic model the jet power is dominated by cold protons, for which we assumed a number of $1\%$ of the electrons. They also dominate the ram pressure.\footnote{Note that the ram pressure only includes the particle kinetic energies.} In the hadronic case, the strong magnetic field dominates the jet power, while the dominance of the protons in the ram pressure increases further through their kinetic energy. In fact, the relative number of protons compared to electrons is smaller in the hadronic case than in the leptonic case \cite{zea17,zea19}. 
The Eddington luminosity of CTA 102's black hole is $P_{\rm edd}\p = 1.1\times 10^{47}\,$erg/s implying that the hadronic model is strongly super-Eddington at all times, a common issue of hadronic models \cite{zb15}. While super-Eddingtion accretion modes are possible over short time-scales \cite{tnm11} and can therefore result in short flares being super-Eddington \cite{kea13,Hea19}, this is probably not realized for the ground state of the jet.

Obviously, both models obtain excellent fits of the long-term lightcurve. The leptonic model requires the efficient scattering of BLR photons, while at the same time the absorption by the BLR should be low \cite{zea19}. Hence, the emission region of the leptonic model is placed at the outer edge of the BLR. As the hadronic model is much less dependent on the soft external photon fields, it has a larger freedom of the location of the emission region. The example shown in Fig.~\ref{fig:had} is placed $1\,$pc from the black hole, outside the BLR (in \cite{zea19} we provide other examples, as well).

From the modeling results, one can infer the cloud parameters, which are also listed in Tab.~\ref{tab:param}. From the changes of the injection luminosity, the number of particles is inferred. The half-duration of the flare (about $60\,$d) provides the radius of the cloud under the assumption that the cloud is on a Keplerian orbit around the black hole. With these values the density in the cloud can be computed. As the scale height, which is a free parameter in our modeling, depends both on the density and the temperature, the latter can be determined. Clearly, the temperatures inferred from the modeling are not reasonable parameters of clouds in close proximity of a hot accretion disk. However, we note that the density is derived under the assumption that all cloud particles are injected into the jet and are accelerated to relativistic speeds. As is known from supernova remnants, the particle acceleration efficiency at a shock is on the order of $10\%$ or less. Additionally, the collision of the cloud with the jet will result in a large fraction of particles being expelled from the collision (by the shock moving through the cloud) instead of being injected into the jet. While this cannot be well quantified, it is probable that actually a large fraction of the cloud is lost and only a small fraction enters the jet. Due to statistical reasons, this still follows the density structure of the cloud keeping Eq.~(\ref{eq:Ninj}) applicable. Hence, the densities listed in Tab.~\ref{tab:param} should be much higher, and in turn would be the temperature.

Nonetheless, it is instructive to speculate on the potential cloud type. While BLR clouds seem to be obvious candidates, they are typically orders of magnitude smaller than what has been inferred here, which makes them unlikely candidates. A large star-forming region is an interesting possibility, as it could also explain the 2-year-symmetry in \cta's lightcurve surrounding the big flare, c.f. Fig.~\ref{fig:ltlc}. The big flare, in turn, would then be the result of a dense hot core (or a proto-star) entering the jet. As blazars are typically hosted by elliptical galaxies, one may ask where such a large cloud could come from, as elliptical galaxies are usually devoid of gas and dust. However, they also are the product of galactic mergers and some of the gas may be hurdled towards the galactic center to fuel the active nucleus. The cloud could be part of such an inflowing stream of gas on its way towards the black hole but that was intercepted by the jet.

If the emission region is instead much further out, where the cloud would be moving even slower, the parameters could also fit well with the atmosphere of a red giant star. The red giant would need to move with $\sim 100\,$km/s, which is a typical stellar speed in elliptical galaxies\footnote{Note that in the models of \cite{bpb12,bea12,bba12} the red giant is located close to the base of the jet and moving a lot faster. In their models, the interaction results in fast flares lasting only a few days, while in our application we obtain a long-lasting flare covering more than 4 months.}. 

In conclusion, the ablation of a slow and/or large cloud by the relativistic jet is a probable explanation of the giant flare observed in \cta, where fluxes rose and fell by factors $10$-$100$ over the course of 4 months.

\section*{Acknowledgement}
The authors wish to thank Maxim Barkov and Valent\'i Bosch-Ramon for stimulating discussions, and the organizers for a wonderful conference.

M.~Z. gratefully acknowledges funding by the German Ministry for Education and Research (BMBF) through grant 05A17PC3.
The work of M.~B. is supported through the South African Research Chair Initiative (SARChI) of the South African Department of Science and Technology (DST) and National Research Foundation.\footnote{Any opinion, finding and conclusion or recommendation expressed in this material is that of the authors, and the NRF does not accept any liability in this regard.}
F.~J. and S.J.~W. acknowledge support by the German Ministry for Education and Research (BMBF) through Verbundforschung Astroteilchenphysik grant 05A11VH2.
J.-P.~L. gratefully acknowledges CC-IN2P3 (\href{https://cc.in2p3.fr}{cc.in2p3.fr}) for providing a significant amount of the computing resources and services needed for this work.
A.~W. is supported by the Foundation for Polish Science (FNP).


\end{document}